\documentclass[aps,prb,onecolumn,10pt,groupedaddress,nofootinbib]{revtex4}
\usepackage[utf8x]{inputenc}
\usepackage{amsmath,amssymb}
\usepackage[dvips]{graphicx}
\begin{document}

\title{Physical observables of the Ising spin glass in $6-\epsilon$ dimensions: asymptotical behavior around the critical
fixed point}

\author{T.~Temesv\'ari}\email{temtam@helios.elte.hu}
\affiliation{MTA-ELTE Theoretical Physics Research Group,
E\"otv\"os University, P\'azm\'any P\'eter s\'et\'any 1/A,
H-1117 Budapest, Hungary
}

\date{\today}

\begin{abstract}
 The asymptotical behavior of physical quantities, like the order parameter, the replicon and longitudinal masses, is studied
 around the zero-field spin glass transition point when a small external magnetic field is applied. An
 effective field theory to model this asymptotics contains a small perturbation in its Lagrangian which breaks the zero-field
 symmetry. A first order renormalization group supplemented by perturbational results provides the scaling functions.
 The perturbative zero of the scaling function for the replicon mass defines a generic Almeida-Thouless surface
 stemming from the zero-field fixed point.
\end{abstract} 


\maketitle

\section{Introduction}
\label{sec1}

Since the invention of the renormalization group (RG) by Wilson
\cite{Wilson_Kogut},
replacing a statistical system (which is close to its critical state) by an effective field theory has become the basic analytical
tool to calculate the asymptotical behavior of physical quantities around a critical point. Such an effective theory is defined
by its Lagrangian $\mathcal L$,
usually called the Landau-Ginzburg-Wilson (LGW) Lagrangian, which depends on the fluctuating order parameter
components, the ``fields'', the statistical weight of a configuration being $\sim e^{-\mathcal L}$.
This formalism has been set up
in the seventies of the last century for the prototype spin glass model of Edwards and Anderson (EA) \cite{EA}, with an immediate application
of the renormalization group \cite{HaLuCh76}. The EA model for $N$ Ising spins on a $d$-dimensional hypercubic lattice is defined by the
Hamiltonian
\begin{equation}\label{EA}
\mathcal{H}=-\sum_{(ij)}J_{ij} s_is_j 
-H\sum_i s_i
\end{equation}
where the $J_{ij}$'s are independent, Gaussian distributed
random variables with zero mean and variance $J^2$,
and a homogeneous external magnetic field $H$ was also included.
Summations are
over nearest neighbour pairs $(ij)$ of lattice sites in the first sum, while over
the $N$ lattice sites in the second one. Averages over the quenched disorder of the EA model are managed by the replica trick, and, as a result,
the effective theory representing the lattice system close to criticality is a cubic {\em replicated\/} field theory with the fluctuating
fields (in momentum space) $\phi^{\alpha\beta}_{\mathbf p}=\phi^{\beta\alpha}_{\mathbf p}$ and $\phi^{\alpha\alpha}_{\mathbf p}=0$ for
$\alpha,\beta=1\dots n$, with the replica number $n$ going to zero at the end of a calculation. Harris et. al.\cite{HaLuCh76}, and later
Refs.\ \cite{PytteRudnick79,BrMo79} too, deduced the following LGW Lagrangian for the zero-external-field case, i.e.\ for $H=0$:
\begin{equation}\label{zero-field}
\mathcal L_{\text{zero-field}}=\frac{1}{2}\sum_{\mathbf p}
\Big(\frac{1}{2} p^2+m\Big)\sum_{\alpha\beta}
\phi^{\alpha\beta}_{\mathbf p}\phi^{\alpha\beta}_{-\mathbf p}
-\frac{1}{6\sqrt{N}}
\sideset{}{'}\sum_{\mathbf {p_1p_2p_3}}
w\sum_{\alpha\beta\gamma}\phi^{\alpha\beta}
_{\mathbf p_1}\phi^{\beta\gamma}_{\mathbf p_2}
\phi^{\gamma\alpha}_{\mathbf p_3}\,\,.
\end{equation}
Momentum conservation is indicated by the primed sum, and a continuum of ${\mathbf p}$'s, cutoff at some $\Lambda$, results in the thermodynamic
limit $N\to \infty$. Replica summations above and in the followings are unrestricted. For a nonzero magnetic field $H$ which is not necessarily
small, the Lagrangian $\mathcal L$ gets additional replica symmetric (RS) invariants
(i.e.\ homogeneous polynomials built up of the fields $\phi^{\alpha\beta}_{\mathbf p}$'s which are invariant under {\em any\/} permutation of the
$n$ replicas),
see Ref.\ \cite{rscikk}, and the theory becomes the generic
cubic RS field theory with $\mathcal L=\mathcal L_{\text{zero-field}}+\delta\mathcal L$, with
$m$ and $w$ in $\mathcal L_{\text{zero-field}}$ replaced by $m_1=m+\delta m_1$ and $w_1=w+\delta w_1$, respectively, and
\begin{multline}\label{delta_L}
\delta\mathcal L=
-\frac{1}{2}N^{\frac{1}{2}}\,h^2\,\sum_{\alpha\beta}
\phi^{\alpha\beta}_{{\mathbf p}=0}+
\frac{1}{2}\,\sum_{\mathbf p}
\bigg[m_2\sum_{\alpha\beta\gamma}
\phi^{\alpha\gamma}_{\mathbf p}\phi^{\beta\gamma}_{-\mathbf p}+
m_3\sum_{\alpha\beta\gamma\delta}
\phi^{\alpha\beta}
_{\mathbf p}\phi^{\gamma\delta}_{-\mathbf p}\bigg]
\\
-\frac{1}{6\sqrt{N}}
\sideset{}{'}\sum_{\mathbf {p_1p_2p_3}}
\bigg[
w_2\sum_{\alpha\beta}\phi^{\alpha\beta}
_{\mathbf p_1}\phi^{\alpha\beta}
_{\mathbf p_2}\phi^{\alpha\beta}_{\mathbf p_3}
+w_3\sum_{\alpha\beta\gamma}\phi^{\alpha\beta}
_{\mathbf p_1}\phi^{\alpha\beta}
_{\mathbf p_2}\phi^{\alpha\gamma}_{\mathbf p_3}
+w_4\sum_{\alpha\beta\gamma\delta}\phi^{\alpha\beta}
_{\mathbf p_1}\phi^{\alpha\beta}
_{\mathbf p_2}\phi^{\gamma\delta}_{\mathbf p_3}
+w_5\sum_{\alpha\beta\gamma\delta}\phi^{\alpha\beta}
_{\mathbf p_1}\phi^{\alpha\gamma}
_{\mathbf p_2}\phi^{\beta\delta}_{\mathbf p_3}\\
+w_6\sum_{\alpha\beta\gamma\delta}\phi^{\alpha\beta}
_{\mathbf p_1}\phi^{\alpha\gamma}
_{\mathbf p_2}\phi^{\alpha\delta}_{\mathbf p_3}
+w_7\sum_{\alpha\beta\gamma\delta\mu}\phi^{\alpha\gamma}
_{\mathbf p_1}\phi^{\beta\gamma}
_{\mathbf p_2}\phi^{\delta\mu}_{\mathbf p_3}
+w_8\sum_{\alpha\beta\gamma\delta\mu\nu}\phi^{\alpha\beta}
_{\mathbf p_1}\phi^{\gamma\delta}
_{\mathbf p_2}\phi^{\mu\nu}_{\mathbf p_3}
\bigg].
\end{multline}
The zero-field Lagrangian of Eq.\ (\ref{zero-field}) contains RS invariants with all the replica indices occuring an even number of times,
thus reflecting the spin inversion symmetry of the EA model without an external magnetic field. Although the insertion of a small magnetic field
breaks the spin inversion symmetry, and consequently the higher symmetry of the field theory with $\mathcal L_{\text{zero-field}}$, the
generic RS field theory with all the coupling constants nonzero in (\ref{delta_L}) is redundant when the magnetic field is small.
Accordingly, the first study of the Almeida-Thouless\cite{AT} (AT) instability below 8 dimensions considered the simplest model with
$h^2$ the only nonzero coupling in (\ref{delta_L})\cite{GrMoBr83}.

Finding the $H$-dependence of the couplings in Eq.\ (\ref{delta_L}) and, in this way, selecting the dominant couplings for small $H$
can be accomplished by two different methods: The first one applies the Gaussian-integral-representation (Hubbard transformation) of the
original EA model, plus additional truncation for small momentum; this was the approach in \cite{rscikk} to derive the generic RS field
theory. Here the second method is chosen, namely that neglecting the fluctuations of the order parameter fields provides the
field-theoretic representation of the {\em infinite\/}-dimensional EA model which is most easily realized on the complete graph
providing the Sherrington-Kirkpatrick (SK) model\cite{SK}; see also Ref.\ \cite{AT2008}.
In the following paragraphs, therefore, the Landau free energy of the field theory is compared with the Lagrangian of the SK model,
$\mathcal L_{\text{SK}}$, which has an explicit and well-known magnetic field dependence.

The SK model has the Hamiltonian (\ref{EA}) on the complete graph, i.e.\ $\sum_{(ij)}$ means summation over all the pairs, and
the variance of the $J_{ij}$'s is $J^2/N$. With the notation of $E[\dots]$ for the average over the $J_{ij}$'s, the quenched
averaged replicated partition function of the SK model can be put into the form
\[
E[Z_{\text{}SK}^n]\sim \int \Big[\prod_{(\alpha\beta)}dq_{\alpha\beta}\Big]\,e^{-\mathcal L_{\text{SK}}}
\]
with
\begin{multline}\label{SK}
\dfrac{1}{N}\mathcal L_{\text{SK}}=-\dfrac{1}{2}\,{\bar H}^2\,\sum_{\alpha\beta}q_{\alpha\beta}
+\dfrac{1}{2}\,(-\bar\tau+{\bar H}^2)\,\sum_{\alpha\beta}
q_{\alpha\beta}^2-\dfrac{1}{2}\,{\bar H}^2\,\sum_{\alpha\beta\gamma}
q_{\alpha\gamma}q_{\beta\gamma}
-\dfrac{1}{6}\,(1-3\,{\bar H}^2)\,\sum_{\alpha\beta\gamma}q_{\alpha\beta}q_{\beta\gamma}q_{\gamma\alpha}
-\dfrac{1}{3}\,{\bar H}^2\,\sum_{\alpha\beta}q_{\alpha\beta}^3\\
+{\bar H}^2\,\sum_{\alpha\beta\gamma}q_{\alpha\beta}^2q_{\beta\gamma}
-\dfrac{1}{2}\,{\bar H}^2\,\sum_{\alpha\beta\gamma\delta}q_{\alpha\beta}q_{\alpha\gamma}q_{\beta\delta}
+O({\bar H}^4,q^4)
\end{multline}
where $\bar\tau\equiv \frac{1}{2}[1-(J/kT)^{-2}]$ and $\bar H\equiv H/kT $.
Stationarity of $\mathcal L_{\text{SK}}$ with respect to $q_{\alpha\beta}$ yields the order parameter in the thermodynamic limit.

In the case of the field theory, it is the Legendre-transformed free energy $F(q_{\alpha\beta})$ that is stationary in the equilibrium
state. It is defined by the common rules of the Legendre transformation, namely
\[
F(q_{\alpha\beta})=-\ln Z(H_{\alpha\beta})+N\,\sum_{(\alpha\beta)}H_{\alpha\beta}\,q_{\alpha\beta}\qquad\text{and}\qquad
\dfrac{\partial\ln Z(H_{\alpha\beta})}{\partial H_{\alpha\beta}}=N\,q_{\alpha\beta}
\]
where the partition function $Z(H_{\alpha\beta})=\int \mathcal D \phi\,e^{-\mathcal L}$ acquires its dependence on the
$H_{\alpha\beta}$'s by adding a source term $-N^{\frac{1}{2}}\,\sum_{(\alpha\beta)}H_{\alpha\beta}\,
\phi^{\alpha\beta}_{{\mathbf p}=0}$ to the RS Lagrangian $\mathcal L_{\text{zero-field}}+\delta\mathcal L$. ($\sum_{(\alpha\beta)}$
in these formulas means summation over the $n(n-1)/2$ pairs of replicas.)
Neglecting fluctuations of the fields (tree approximation), i.e.\  replacing $\phi^{\alpha\beta}_{{\mathbf p}}$ by its
average $\langle\phi^{\alpha\beta}_{{\mathbf p}}\rangle=\delta^{\text{Kr}}_{\mathbf p,\mathbf 0}\times N^{\frac{1}{2}}\,
q_{\alpha\beta}$, provides the mean field, or Landau, free energy of the model:
\begin{multline}\label{F}
\dfrac{1}{N}F(q_{\alpha\beta})=-\dfrac{1}{2}\,h^2\sum_{\alpha\beta}q_{\alpha\beta}
+\dfrac{1}{2}\, \Big[(m+\delta m_1)\sum_{\alpha\beta}q_{\alpha\beta}^2+m_2\sum_{\alpha\beta\gamma}q_{\alpha\gamma}q_{\beta\gamma}
+m_3\sum_{\alpha\beta\gamma\delta}q_{\alpha\beta}q_{\gamma\delta}\Big]\\
-\dfrac{1}{6}\,\Big[(w+\delta w_1)\sum_{\alpha\beta\gamma}q_{\alpha\beta}q_{\beta\gamma}q_{\gamma\alpha}
+w_2\sum_{\alpha\beta}q_{\alpha\beta}^3+w_3\sum_{\alpha\beta\gamma}q_{\alpha\beta}^2q_{\beta\gamma}
+w_4\sum_{\alpha\beta\gamma\delta}q_{\alpha\beta}^2q_{\gamma\delta}+w_5\sum_{\alpha\beta\gamma\delta}q_{\alpha\beta}q_{\alpha\gamma}q_{\beta\delta}\\
+w_6\sum_{\alpha\beta\gamma\delta}q_{\alpha\beta}q_{\alpha\gamma}q_{\alpha\delta}+w_7\!\!\sum_{\alpha\beta\gamma\delta\mu}q_{\alpha\gamma}q_{\beta\gamma}
q_{\delta\mu}+w_8\!\!\!\sum_{\alpha\beta\gamma\delta\mu\nu}q_{\alpha\beta}q_{\gamma\delta}q_{\mu\nu}\Big]+O(q^4)\,.
\end{multline}
Comparing Eqs.\ (\ref{SK}) and (\ref{F}), one can conclude for the bare couplings of the field theory:
\begin{itemize}
 \item Writing $m\equiv m_c-\tau$ with $\tau= 0$ at the critical point of the field theory, $m_c\sim (T_c^2-{T_c^{\text{mf}}}^2)$ results where
 $T_c$ and $T_c^{\text{mf}}$ are the critical temperatures of the field theory and its mean field approximation, respectively. This shows that
 $m_c$ is one-loop order.
 \item The couplings $h^2$, $\delta m_1$, $m_2$, $\delta w_1$, $w_2$, $w_3$, and $w_5$ are of order $ {\bar H}^2$, whereas all
 the other couplings are at most of order ${\bar H}^4$.
\end{itemize}

A simple three-parameter model was used in Ref.\ \cite{PT} to study, among other things, the AT instability for $6<d<8$, $d=6$, and
$d\lesssim 6$,
the nonzero bare parameters were $m_1=m= m_c-\tau$, $w_1= w$, and $h^2$.
It was found in \cite{PT} that the critical field $h^2_{\text{AT}}$ behaves continuously while crossing the upper critical dimension 6
for fixed values of the reduced-temperature-like parameter $\tau$ and cubic coupling $w$,
and the AT line takes the simple form
\begin{equation}\label{AT_d=6}
h^2_{\text{AT}}\approx
\dfrac{4}{(1-w^2\ln \tau)^4}\,\, w\,\tau^2, \qquad\qquad d=6,
\end{equation}
valid if $\tau\ll 1$ and $w^2\ll 1$, in exactly the upper critical dimension.
As the main motivation of the present paper, we want to check whether a suitable extension of this simple model 
in such a way that, beside $h^2$, the bare couplings $m_2$, $w_2$, $w_3$, and $w_5$ are small but nonzero too,
will or will not
modify the results of Ref.\ \cite{PT} about the AT instability around the zero-field critical point, and for $d\lesssim 6$.
In the dimensional regime $6<d<8$ where a standard perturbational method is applicable, an extended parameter space with all the
couplings which are of order ${\bar H}^2$ seems to be a convenient extension. Below six dimensions, however, where the simple perturbative
method breaks down (due to the more and more infrared divergent graphs as the number of loops increases), it becomes inevitable to apply
the RG for the calculation of the asymptotical behavior of physical quantities. In this case, however, it is difficult to define
the model by the set of bare couplings (by those, for instance, which are at least of order ${\bar H}^2$), as new couplings will be
generated by the RG flow.

In the present paper, we propose to define the model by the set of nonlinear scaling fields: this ensures the closedness of the model
under RG flow. The simple three-parameter model of Ref.\ \cite{PT} can be formulated in this way, and its extension will be done by
introducing a new (mass-like) nonlinear scaling field which, on the level of the bare couplings, leads to a more complicated model.
In this more complicated field theory, one can calculate in $6-\epsilon$ dimensions
the RS order parameter, the replicon and longitudinal masses; all in the framework of first order RG
combined with perturbational analysis. We focus on the asymptotical behavior close to the zero-field critical fixed point.
The perturbative zero of the replicon mass defines the onset of the instability of the RS phase (AT surface).
The problem of the runaway RG flows along this partially massless, i.e.\ massless in the replicon sector, manifold (caused by the repulsion of the critical
fixed point) is also discussed.

The outline of the paper is as follows: The method of using nonlinear scaling fields for the calculation of physical quantities
below 6 dimensions is discussed in Section \ref{sec2}. The results in this section are equivally valid below and above the critical
temperature. The free propagators (replicon and longitudinal) are constructed in Section \ref{sec3}.
The central part of the paper is Section \ref{sec4} where the critical asymptotics of physical quantities, such as
the order parameter, the replicon and longitudinal masses, are elaborated. The more interesting case of $T<T_c$ is presented
in subsection \ref{below}, whereas results for $T>T_c$ are also displayed for the sake of completeness and comparison in \ref{above}.
The limitations of the various approximations used to achieve our results are discussed in some details in the next section.
Zeros of the replicon mass are found in a region of the parameter space around the critical fixed point which belongs to
the range of applicability of our approximations. There is also a discussion of this Almeida-Thouless critical manifold
in Section \ref{D}. Some conclusive remarks and a paragraph about the applied perturbative method are left to Section \ref{conclusion}.
The basic perturbative formulas are displayed in the Appendix.

Many results in this paper, especially the connection between bare parameters and nonlinear scaling fields in Section \ref{sec2},
are built upon the first order RG equations of Ref.\ \cite{Iveta}.

\section{Below 6 dimensions}
\label{sec2}

The RG equations for the generic cubic field theory defined in Eqs.\ (\ref{zero-field}) and  (\ref{delta_L}) can be obtained
by integrating out degrees of freedom in a momentum shell at the cutoff $\Lambda$. The structure of these flow equations in the
one-loop approximation, and for $n=0$, can be
written as:
\begin{equation}\label{RG_flow}
\begin{aligned}
 \dot{h^2}&=\Big[4-\dfrac{\epsilon}{2}-{\mathcal H}^{(2)}(m_1,m_2,m_3;w_1,\dots,w_8)\Big]\,h^2+{\mathcal H}^{(1)}(m_1,m_2,m_3;w_1,\dots,w_8);\\[4pt]
 \dot m_i&=2\,m_i+{\mathcal M}_i^{(2)}(m_1,m_2,m_3;w_1,\dots,w_8),\,\,\quad\, i=1,2,3;\\[4pt]
 \dot w_i&=\dfrac{\epsilon}{2}\,w_i+{\mathcal W}_i^{(3)}(m_1,m_2,m_3;w_1,\dots,w_8),\quad\,\,\,\, i=1,\dots,8.
\end{aligned}
\end{equation}
The functions ${\mathcal F}^{(k)}(m_1,m_2,m_3;w_1,\dots,w_8)$ above (with $\mathcal F=\mathcal H$, ${\mathcal M}_i$, or ${\mathcal W}_i$)
are homogeneous polynomials of degree $k$ in the $w$'s, while analytic in the masses with a nonzero value for $m_1=m_2=m_3=0$.
All but the first equations in (\ref{RG_flow})
has been published in Ref.\ \cite{Iveta}, although the set of bare couplings was chosen differently
there.\footnote{For the sake of easing the reader, we give here the precise citations where the linear connection between the two sets of couplings
can be found: For the masses (i.e.\ $m_R$, $m_A$, and $m_L$ in \cite{Iveta} versus $m_1$, $m_2$, and $m_3$ here) see Eqs.~(32) of \cite{Iveta} and
Eqs.~(22-24) of \cite{rscikk}, whereas for the cubic couplings (i.e.\ $g_i$, $i=1\dots 8$ in \cite{Iveta} versus
$w_i$, $i=1\dots 8$ here) see Eqs.~(49a-h) of \cite{rscikk}.} The flow equation for the magnetic field in the generic case, however, has not been published
before:
\begin{equation}\label{h^2}
\dot{h^2}=\left(4-\frac{\epsilon}{2}-\frac{1}{2}\,\eta_L\right)\,h^2+(3g_3+3g_6+2\bar g_7)\,\frac{1}{1+2m_1}
+(3g_6+2\bar g_7)\,\frac{2m_2}{(1+2m_1)(1+2m_1-2m_2)}-2g_6\,\frac{m_2-2m_3}{(1+2m_1-2m_2)^2}\,,
\end{equation}
with
\begin{equation}\label{etaL}
 \eta_L=2g_3^2\,\frac{1+6m_1}{(1+2m_1)^4}-\frac{8}{3}\,(g_6^2+g_6\bar g_7)\,\frac{1+6m_1-6m_2}{(1+2m_1-2m_2)^4}
+\frac{4}{3}\,g_6^2\,(m_2-2m_3)\,\frac{1+18m_1-18m_2}{(1+2m_1-2m_2)^5}
 \end{equation}
 
where we adopted the notations from Ref.~\cite{Iveta} \footnote{\label{mistake}The last term in $\eta_L=\eta_A$ is wrongly missing in Eq.\ (87) of Ref.~\cite{Iveta}.
A similar term proportional to $m_2-2m_3$ was also left out from the expression Eq.\ (86) for $\eta_R$.}:
\begin{align*}
 g_3&\equiv -w_1+w_2-\frac{1}{3}\,w_3,\\
 g_6&\equiv 2w_1-w_2+w_3-w_5-w_6,\\
 \bar g_7&\equiv -\frac{3}{2}\,w_1+\frac{1}{2}\,w_2-\frac{5}{6}\,w_3+\frac{2}{3}\,w_4+\frac{4}{3}\,w_5+w_6-\frac{2}{3}\,w_7\,.
\end{align*}

One can benefit the following information from the RG equations (\ref{RG_flow}):
\begin{itemize}
 \item The zeros of the right-hand-side provide the fixed points. In this paper, we are interested in the vicinity of the zero-field critical
 fixed point: ${w^*}^2=\frac{1}{2}\,\epsilon$, $m^*=-\frac{1}{2}\,{w^*}^2=-\frac{1}{4}\,\epsilon$, and all the other couplings being zero.
 We prefer using $2{w^*}^2$, instead of $\epsilon$, in the remainder part of the paper.
 \item All the eigenmodes of the linearized RG equations, with the only exception of that belonging to $h^2$, were published in \cite{Iveta}.
 In this paper we restrict ourself to a model with the following four modes:
 \begin{equation}\label{modes}
\begin{gathered}
g_{h^2}\quad\text{with}\quad\lambda_{h^2}=4-\frac{2}{3}\,{w^*}^2,\qquad g_{m_1}\quad\text{with}\quad\lambda_{m_1}=2-\frac{10}{3}\,{w^*}^2,\qquad
g_{m_2}\quad\text{with}\quad\lambda_{m_2}=2-\frac{4}{3}\,{w^*}^2,\\[6pt]
\text{and}\quad g_{w}\quad\text{with}\quad\lambda_{w}=-2{w^*}^2\,\,.
\end{gathered}
\end{equation}
\item The $g$'s above, with subscripts $h^2$, $m_1$, $m_2$, and $w$ referring the modes they belong to, are nonlinear scaling fields%
\footnote{The general theory of the application of nonlinear scaling
fields was briefly summarized in Sec.~5.1 of Ref.~\cite{PT}. The concept of nonlinear scaling fields was introduced by Wegner \cite{Wegner}.}
which 
satisfy exactly, by definition, the equations $\dot{g}=\lambda_{}\, g_{}$ and are zero at the fixed point.
By means of the RG equations (\ref{RG_flow}) above, one can express the original bare couplings in terms of the $g$'s.
Keeping the fields
which break the zero-field symmetry (i.e.\ $g_{h^2}$ and $g_{m_2}$) linear in these expressions (which is sufficient for a small
external field), only the following couplings in $\delta \mathcal L$ are generated:
\begin{equation}\label{bare1}
\begin{aligned}
w^*h^2&=\left(1-\frac{1}{3}\,g_{w}-\frac{1}{3}\,{w^*}^2g_{m_1}\right)\,g_{h^2}+\left(-{w^*}^2-\frac{7}{3}\,{w^*}^2\,g_{w}
+2\,g_{m_1}\right)\,g_{m_2}\,,\\[8pt]
m_2&=\left(1+\frac{4}{3}\,g_{w}+5 {w^*}^2g_{m_1}  \right)\,g_{m_2}\,,\\[8pt]
w_2/w^*&=\left(-12 {w^*}^2-52\,{w^*}^2 \,g_{w}+48\, {w^*}^2g_{m_1}\right)\,g_{m_2}\,\,,\\[8pt]
w_3/w^*&=\left(\frac{49}{2}\,{w^*}^2+\frac{637}{6}\,{w^*}^2 \,g_{w}-94\, {w^*}^2g_{m_1}\right)\,g_{m_2}\,\,,\\[8pt]
w_4/w^*&=\left(-\frac{9}{2}\,{w^*}^2-\frac{39}{2}\,{w^*}^2 \,g_{w}+18\, {w^*}^2g_{m_1}\right)\,g_{m_2}\,\,,\\[8pt]
w_5/w^*&=\left(-\frac{1}{2}\,{w^*}^2-\frac{13}{6}\,{w^*}^2 \,g_{w}-2 {w^*}^2g_{m_1}\right)\,g_{m_2}\,\,;
\end{aligned}
\end{equation}
whereas the symmetric couplings $m_1$ and $w_1$ are
\begin{equation}\label{bare2}
\begin{aligned}
 m_1-m^*&= \left[g_{m_1}-{w^*}^2 \,g_{w}+\frac{10}{3}\, g_{m_1} \,g_{w}-2{w^*}^2\,g_{w}^2
 +\frac{16}{3}\,{w^*}^2\,g_{m_1}^2\right]
+ \left(-1-\frac{4}{3}\,g_{w}-5 {w^*}^2g_{m_1}\right)\,g_{m_2}\,\,,\\[10pt]
w_1/w^*-1&=\left[ 5{w^*}^2g_{m_1}+g_{w}+\frac{190}{6}\,{w^*}^2\, g_{m_1} \,g_{w}+\frac{3}{2}\,g_{w}^2
 -14\,{w^*}^2\,g_{m_1}^2\right]\\[7pt]
&\quad\,+ \left(\frac{1}{2}\, {w^*}^2+\frac{13}{6}\,{w^*}^2 \,g_{w}+2\, {w^*}^2g_{m_1}\right)\,g_{m_2}\,\,.
\end{aligned}
\end{equation}
(The zero-field-symmetric part above has been written up to quadratic order in $g_{m_1}$ and $g_{w}$.)
\end{itemize}

The three-parameter model of Ref.\ \cite{PT} corresponds to the three scaling fields: $g_{m_1}$ and $g_{w}$ span the symmetric (zero-field)
system, whereas $g_{h^2}$ breaks this symmetry. Having a look at Eqs.\ (\ref{bare1}) and (\ref{bare2}), one can realize that $h^2$ is the 
only coupling of the symmetry breaking part $\delta\mathcal L$ which is generated. Therefore, this model can be equivalently defined
by the bare couplings $m_1=m$, $w_1=w$, and $h^2$.

In the present paper, we supplement the model by $g_{m_2}$, whose introduction considerably complicates the model when it is written as
in Eqs.\ (\ref{zero-field}) and  (\ref{delta_L}).
(One cannot avoid using this representation when, for instance, a scaling function is to be calculated.)
The following couplings enter for a small $g_{m_2}$, according to Eqs.\ (\ref{bare1}) and (\ref{bare2}):
$\delta m_1$, $\delta w_1$, $m_2$, $w_2$, $w_3$, $w_4$, and $w_5$.

Any observable $\mathcal O$ can now be considered as depending on the four scaling fields, and according to the generic theory
in Sec.\ 5.1 of \cite{PT}, one can write the following {\em asymptotically exact\/} expression around the fixed point:
\begin{multline}\label{generic_scaling}
\mathcal O(g_{m_1},g_{w};g_{h^2},g_{m_2})=|g_{m_1}|^{\frac{k}{\lambda_{m_1}}}
\,\,\,\hat{\mathcal O}(x,y,z)\\[8pt]
\times \left[1+\frac{k_{m_1}}{\lambda_{m_1}}g_{m_1}
+ \frac{k_{w}}{\lambda_{w}}\,|g_{m_1}|^{\frac{\lambda_{w}}{\lambda_{m_1}}}\,x 
+ \frac{k_{h^2}}{\lambda_{h^2}}\,|g_{m_1}|^{\frac{\lambda_{h^2}}{\lambda_{m_1}}}\,y 
+\frac{k_{m_2}}{\lambda_{m_2}}\,|g_{m_1}|^{\frac{\lambda_{m_2}}{\lambda_{m_1}}}\, z
+\dots 
\right]
\end{multline}
where the RG {\em invariants\/} are defined as
\begin{equation}\label{invariants}
x\equiv  g_{w}\,|g_{m_1}|^{-\frac{\lambda_{w}}{\lambda_{m_1}}}\,,\qquad
y\equiv g_{h^2}\,|g_{m_1}|^{-\frac{\lambda_{h^2}}{\lambda_{m_1}}}\,,\qquad 
z\equiv g_{m_2}\,|g_{m_1}|^{-\frac{\lambda_{m_2}}{\lambda_{m_1}}}\,.
\end{equation}
The \dots\  symbol means neglected terms, namely higher powers of the temperature-like field $g_{m_1}$ and/or
quadratic or higher order monomials of the RG invariants. The $k$'s above are defined for a given $\mathcal O$ by
the RG flow of it as
\begin{equation}\label{dotO}
\dot{\mathcal O}=(k+k_{m_1}g_{m_1}+k_{w}g_{w}+k_{h^2}g_{h^2}+k_{m_2}g_{m_2}+\dots)\,\mathcal O\,.
\end{equation}
The scaling function $\hat{\mathcal O}$ is not determined by the renormalization group, but auxiliary information
is needed  (perturbative method, for instance) to compute it.
Hereinafter we study three observables: the RS order parameter $q$, the replicon and longitudinal masses, i.e.\ 
$\Gamma_R$ and $\Gamma_L$.

\section{Free propagators of the model}
\label{sec3}

When the order parameter $q$ is nonzero, a reorganization of the perturbational series by the shift
$\phi^{\alpha\beta}_{\mathbf p}\rightarrow\phi^{\alpha\beta}_{\mathbf p}
-\sqrt{N}\,q\,\delta^{\text{Kr}}_{\mathbf p=0} $ of the fluctuating fields is useful, as one gets then rid of
``tadpole'' insertions. As a result, the bare magnetic field and the masses suffer similar shifts:
\begin{align*}
h^2\rightarrow\bar{h^2}&= h^2+(-2m_1+2m_2)\,q+(-2w_1+w_2-w_3+w_5)\,q^2\,,\\[5pt]
m_1\rightarrow\bar{m_1}&=m_1+\left(w_1-w_2+\frac{1}{3}\,w_3\right)\,q\,,\\[5pt]
m_2\rightarrow\bar{m_2}&=m_2+\left(-w_1-\frac{2}{3}\,w_3+w_5\right)\,q\,,\\[5pt]
m_3\rightarrow\bar{m_3}&=m_3+\left(-\frac{2}{3}\,w_4-\frac{1}{3}\,w_5\right)\,q\,.
\end{align*}
In the $n\to 0$ limit, two free propagators emerge in the generic RS theory, namely
\[
\bar G_R=\frac{1}{p^2+2\bar{m_1}},\quad\text{the replicon propagator, and}\quad \bar G_L=\frac{1}{p^2+2\bar{m_1}-2\bar{m_2}},
\quad\text{the longitudinal propagator}.
\]
Any perturbative contribution for some observable will, therefore, depend on $q$ which must be computed
from the equation of state, i.e.\ from the condition $\langle \phi^{\alpha\beta}_{\mathbf p}\rangle=0$.
For the free propagators, we need the tree (zero-loop) approximation of this equation:
\begin{equation*}
 2 w^*q=h^2\,q^{-1}-2m_1+2m_2+\left[ -2(w_1-w^*)+w_2-w_3+w_5\right]\,q\,.
\end{equation*}
Using Eqs.\ (\ref{bare1}) and (\ref{bare2}) together with the definitions of the RG invariants in (\ref{invariants}),
the zero-loop order parameter follows, up to first order in $x$, $y$, and $z$, as:
\[
w^*q=|g_{m_1}|\times
     \begin{cases}
     1+\dfrac{7}{3}x+z+\dfrac{1}{2} y & \qquad\text{if $g_{m_1}<0$, i.e.\ $T<T_c$}\\[15pt]
     z+\dfrac{1}{2} y& \qquad\text{if $g_{m_1}>0$, i.e.\ $T>T_c$.}
     \end{cases}
\]
This is the point where the calculations above and below $T_c$ separate. Writing the free propagators as
 \begin{equation}\label{propagators}
 \bar G_R= \frac{1}{p^2+|g_{m_1}|\times R}\qquad \text{and}\qquad \bar G_L= \frac{1}{p^2+|g_{m_1}|\times L}\,\,,
 \end{equation}
it is obtained in the two respective regimes:
\begin{itemize}
 \item $T<T_c$:
 \begin{equation}\label{<}
  R=y\qquad \text{and}\qquad L=2+\dfrac{20}{3}x+2y\,,\qquad g_{m_1}<0\,;
 \end{equation}
\item $T>T_c$:
 \begin{equation}\label{>}
  R=2+\dfrac{20}{3}x+y\qquad \text{and}\qquad L=2+\dfrac{20}{3}x+2y\,,\qquad g_{m_1}>0\,.
 \end{equation} 
\end{itemize}
Higher than first order terms in $x$, $y$, and $z$ are again neglected in the above formulas, in accordance with the smallness of these RG invariants.

\section{Asymptotical behavior around $T_c$} 
\label{sec4}

\subsection{Below $T_c$ ($g_{m_1}<0$)}\label{below}

\subsubsection{The order parameter $q$}

The RG flow for $q$ is simply $\dot q=(2-{w^*}^2+\eta_L/2)\,q$ with $\eta_L$ in (\ref{etaL}). Inserting the nonlinear
scaling fields by the help of (\ref{bare1}) and (\ref{bare2}), the $k$ coefficients for $q$ can be read off by the general definition
in (\ref{dotO}): $k=2-\frac{4}{3}\,{w^*}^2$, $k_{m_1}=\frac{2}{3}\,{w^*}^2$, $k_{w}=-\frac{2}{3}\,{w^*}^2$,
$k_{h^2}=0$, and $k_{m_2}=-\frac{2}{3}\,{w^*}^2$. Using the eigenvalues of the various modes from
(\ref{modes}), the generic scaling form in (\ref{generic_scaling}) becomes
\begin{equation}\label{scaling_q}
q=|g_{m_1}|^{1+{w^*}^2}
\,\,\,\hat{q}(x,y,z)
\times \left[1+\frac{1}{3}\,{w^*}^2\,g_{m_1}
+ \frac{1}{3}\,|g_{m_1}|^{-{w^*}^2}\,x 
-\frac{1}{3}\,{w^*}^2\,|g_{m_1}|^{1+{w^*}^2}\, z
+\dots 
\right]\,.
\end{equation}
Comparing this RG formula with its perturbative counterpart%
\footnote{The notation $O({w^*}^2)$ will be consistently used in this
section whenever the corresponding correction is not available in this
first order RG calculation.}
\begin{equation}\label{q_pert}
\begin{aligned}
w^* q&=
|g_{m_1}|\left\{\big[1+(2+\ln 2)\,{w^*}^2\big]+\frac{1}{2}\,\big[1+(4-2\ln 2)\,{w^*}^2\big]\,y
 +\frac{7}{3}\,\big[1+O({w^*}^2)\big]\,x+\big[1+O({w^*}^2)\big]\,z
 \right\}\\[6pt]
 &\,\,\,\,\,\,+{w^*}^2\,(|g_{m_1}|\,\ln |g_{m_1}|)\,\Big(1+\frac{1}{2}\,y+2x+z\Big)
 \end{aligned}
\end{equation}
[which follows from Eq.\ (\ref{q}) by using of (\ref{bare1}), (\ref{bare2}), (\ref{modes}), (\ref{invariants}), (\ref{Y}), and (\ref{<})],
has a double use:
Firstly, the scaling function can be derived as
\begin{equation}\label{q_hat}
w^*\hat q =\big[1+(2+\ln 2)\,{w^*}^2\big]+\frac{1}{2}\,\big[1+(4-2\ln 2)\,{w^*}^2\big]\,y
+\big[2+O({w^*}^2)\big]\,x+\big[1+O({w^*}^2)\big]\,z\,.
\end{equation}
Secondly, the logarithm in (\ref{q_pert}) should correctly exponentiate in accordance with
the asymptotic scaling above: this property is easily checked by comparison.

\subsubsection{The replicon mass}

The replicon mass satisfies the equation $\dot{\,\,\,\,\Gamma_R}=(2-\eta_R)\,\Gamma_R$, with $\eta_R$ computed in Ref.~\cite{Iveta}.
(See also footnote \ref{mistake}.) Instead of providing the complete formula for $\eta_R$ here again, we show it expressed and
linearly truncated in terms of the nonlinear scaling fields:
\[
\eta_R=-\frac{2}{3}\,{w^*}^2(1-2g_{m_1}+g_{m_2}+2g_w)\,.
\]
The $k$ coefficients (of $\Gamma_R$) follow then by (\ref{dotO}): $k=2+\frac{2}{3}\,{w^*}^2$, $k_{m_1}=-\frac{4}{3}\,{w^*}^2$,
$k_{w}=\frac{4}{3}\,{w^*}^2$, $k_{h^2}=0$, and $k_{m_2}=\frac{2}{3}\,{w^*}^2$. The generic result (\ref{generic_scaling})
can then be translated to the case of the replicon mass, see also (\ref{modes}), as
\begin{equation}\label{scaling_R}
 \Gamma_R=|g_{m_1}|^{1+2{w^*}^2}
\,\,\,\hat{\Gamma_R\!\!\!}\,\,\,(x,y,z)
\times \left[1-\frac{2}{3}\,{w^*}^2\,g_{m_1}
- \frac{2}{3}\,|g_{m_1}|^{-{w^*}^2}\,x
+\frac{1}{3}\,{w^*}^2\,|g_{m_1}|^{1+{w^*}^2}\, z
+\dots
\right]\,.
\end{equation}
The corresponding perturbative formula follows from (\ref{Gamma_R}) and the use of Eqs.\
(\ref{bare1}), (\ref{bare2}), (\ref{modes}), (\ref{invariants}), (\ref{sigma_R}), and (\ref{<}):
\begin{equation}\label{R_pert}
 \Gamma_R=|g_{m_1}|\left\{-4{w^*}^2+\big[1+(-8+3\ln 2-4\ln y)\,{w^*}^2\big]\,y+O({w^*}^2)\,x
 +O({w^*}^2)\,z
 \right\}+{w^*}^2\,(|g_{m_1}|\,\ln |g_{m_1}|)\,2y\,.
\end{equation}
Matching these two expressions of the replicon mass provides 
the scaling function:
\begin{equation}\label{R_hat}
\hat{\Gamma_R\!\!\!}\,\,\,(x,y,z)=-4{w^*}^2+\big[1+(-8+3\ln 2-4\ln y)\,{w^*}^2\big]\,y+O({w^*}^2)\,x
 +O({w^*}^2)\,z\,,
\end{equation}
and it is easy to check that the criterion of proper exponentiation is satisfied.

\subsubsection{The longitudinal mass}

The $k$ coefficients, defined in (\ref{dotO}), for $\Gamma_L$ follow from its RG equation $\dot{\Gamma}_L=(2-\eta_L)\,\Gamma_L$ and
Eqs.\ (\ref{etaL}), (\ref{bare1}), and (\ref{bare2}): $k=2+\frac{2}{3}\,{w^*}^2$, $k_{m_1}=-\frac{4}{3}\,{w^*}^2$,
$k_{w}=\frac{4}{3}\,{w^*}^2$, $k_{h^2}=0$, and $k_{m_2}=\frac{4}{3}\,{w^*}^2$. Just as for the replicon
case, one can easily conclude the scaling form of the longitudinal mass as
\begin{equation}\label{scaling_L}
 \Gamma_L=|g_{m_1}|^{1+2{w^*}^2}
\,\,\,\hat{\Gamma_L\!\!\!}\,\,\,(x,y,z)
\times \left[1-\frac{2}{3}\,{w^*}^2\,g_{m_1}
- \frac{2}{3}\,|g_{m_1}|^{-{w^*}^2}\,x
+\frac{2}{3}\,{w^*}^2\,|g_{m_1}|^{1+{w^*}^2}\, z
+\dots
\right]\,
\end{equation}
which can be confronted with (\ref{Gamma_L}) of the Appendix:
\begin{equation}\label{L_pert}
\begin{aligned}
\Gamma_L&=|g_{m_1}|\left\{\big[2+(-8+4\ln 2)\,{w^*}^2\big]+\big[2+(1+4\ln 2-6\ln y)\,{w^*}^2\big]\,y
+\frac{20}{3}\,\big[1+O({w^*}^2)\big]\,x+O({w^*}^2)\,z
\right\}\\[6pt]
&\,\,\,\,\,\,+4{w^*}^2\,(|g_{m_1}|\,\ln |g_{m_1}|)\,\Big(1+y+\frac{11}{3}\,x\Big)\,.
\end{aligned}
\end{equation}
[Use of Eqs.\ (\ref{bare1}), (\ref{bare2}), (\ref{modes}), (\ref{invariants}), (\ref{sigma_L}), and (\ref{<})
is necessary to put (\ref{Gamma_L}) into this form.]
The scaling function can now be read off as
\begin{equation}\label{L_hat}
\hat{\Gamma_L\!\!\!}\,\,\,(x,y,z)=\big[2+(-8+4\ln 2)\,{w^*}^2\big]+\big[2+(1+4\ln 2-6\ln y)\,{w^*}^2\big]\,y
+\big[8+O({w^*}^2)\big]\,x+O({w^*}^2)\,z\,,
\end{equation}
and exponentiation can be checked.

\subsection{Results for $T_c>0$ ($g_{m_1}>0$)}\label{above}

For the sake of completeness and a possible comparison with the $T_c<0$ case, results for the three observables above
the critical temperature (in a small but finite magnetic field) are presented in this subsection. Their scaling forms
in Eqs.\ (\ref{scaling_q}), (\ref{scaling_R}), and (\ref{scaling_L}) are equally valid in this high temperature asymptotical
regime, the scaling functions, however, are different. Due to the change of the free propagators according to (\ref{propagators})
and (\ref{>}), the one-loop perturbative results are now [instead of (\ref{q_pert}), (\ref{R_pert}), and (\ref{L_pert})]:
\begin{align*}
w^* q&=
g_{m_1}\left\{\frac{1}{2}\,\big[1-(1+2\ln 2)\,{w^*}^2\big]\,y
 +\big[1+O({w^*}^2)\big]\,z\right\}+
{w^*}^2\,(g_{m_1}\,\ln g_{m_1})\,\Big(\frac{1}{2}\,y+z\Big)\,,\\[10pt]
 \Gamma_R&=g_{m_1}\left\{2\big[1+(1+2\ln 2)\,{w^*}^2\big]+\dfrac{1}{2}\big[2+(1-2\ln 2)\,{w^*}^2\big]\,y
+\Big[\dfrac{20}{3}+O({w^*}^2)\Big]\,x +O({w^*}^2)\,z
\right\}\\[5pt]
&\quad\,+{w^*}^2\,(g_{m_1}\,\ln g_{m_1})\,\Big[4+2y+\dfrac{44}{3}\,x\Big]\,,\\[10pt]
\Gamma_L&=g_{m_1}\left\{2\big[1+(1+2\ln 2)\,{w^*}^2\big]+\dfrac{1}{2}\big[4+(5+2\ln 2)\,{w^*}^2\big]\,y
+\Big[\dfrac{20}{3}+O({w^*}^2)\Big]\,x +O({w^*}^2)\,z
\right\}\\[6pt]
&\quad\,+{w^*}^2\,(g_{m_1}\,\ln g_{m_1})\,\Big[4+4y+\dfrac{44}{3}\,x\Big]\,.
\end{align*}
Comparing with the scaling forms in Eqs.\ (\ref{scaling_q}), (\ref{scaling_R}), and (\ref{scaling_L}), the scaling
functions above the critical temperature can be concluded:
\begin{align}
 w^*\hat q &=\frac{1}{2}\,\big[1-(1+2\ln 2)\,{w^*}^2\big]\,y
+\big[1+O({w^*}^2)\big]\,z\,,\label{q_hat_>}\\[10pt]
\hat{\Gamma_R\!\!\!}\,\,\,(x,y,z)&=2\big[1+(1+2\ln 2)\,{w^*}^2\big]+\big[1+\dfrac{1}{2}\,(1-2\ln 2)\,{w^*}^2\big]\,y+
\big[8+O({w^*}^2\big]\,x
 +O({w^*}^2)\,z\,,\label{R_hat_>}\\[10pt]
 \hat{\Gamma_L\!\!\!}\,\,\,(x,y,z)&=2\big[1+(1+2\ln 2)\,{w^*}^2\big]+\big[2+\dfrac{1}{2}\,(5+2\ln 2)\,{w^*}^2\big]\,y+
\big[8+O({w^*}^2\big]\,x
 +O({w^*}^2)\,z\,.\label{L_hat_>}
\end{align}

One can make the following observations about the behavior of the three quantities around the critical
point:
\begin{itemize}
 \item The high-temperature ($g_{m_1}>0$) and zero-external-magnetic-field ($y=z=0$) phase possesses a higher symmetry
 with zero order parameter [see (\ref{q_hat_>})] and a single mass [due to the degeneration between the replicon
 and longitudinal masses, see Eqs.\ (\ref{R_hat_>}) and (\ref{L_hat_>})].
 \item In zero-external-magnetic-field ($y=z=0$) below the critical temperature ($g_{m_1}<0$), the order parameter is
 nonzero [Eq.\ (\ref{q_hat}): this is the RS spin glass phase invented by Edwards and Anderson \cite{EA}.
 However, according to Eq.\ (\ref{R_hat}), the replicon mass gets negative due to the one-loop term, showing that this phase
 is unstable, just as in mean field theory \cite{SK}, and replica symmetry must be broken.
 \item There is a slight splitting between the replicon and longitudinal masses in a small magnetic field above $T_c$,
 Eqs.\ (\ref{R_hat_>}) and (\ref{L_hat_>}), whereas the longitudinal mass is definetely massive below $T_c$, Eq.\ (\ref{L_hat}),
 and therefore separates from the replicon one.
 \item It is obvious from Eq.\ (\ref{R_hat}) that stability of the RS phase is restored for $y>y_0\sim O({w^*}^2)$ and $g_{m_1}<0$.
\end{itemize}

\section{Discussion: range of applicability and asymptotically detected Almeida-Thouless instability}\label{D}

In deriving our basic results for the scaling forms and scaling functions of the three observables ($q$, $\Gamma_R$, and $\Gamma_L$),
several approximations were applied in the previous section. For seeing clearly the limits of these approximations, it might
be useful to give an overall list of them here:
\begin{itemize}
 \item The RG equations and the auxiliary perturbative calculations have the one-loop character, and therefore ${w^*}^2=\epsilon/2\ll 1$.
 \item The multiplicative factor 
 (which is analytic in the fields $g_{m_1}$,
 $g_w$, $g_{h^2}$, and $g_{m_2}$) in
 the, in principle exact, scaling formula of Eq.\ (\ref{generic_scaling}) was truncated to linear
 order in the nonlinear scaling fields. We must have, therefore, $|g_{m_1}|$, $|g_w|$, $|g_{m_2}|$, and $g_{h^2}$ much
 smaller than unity. In fact, the normalization of the nonlinear scaling fields (which is not fixed originally) was
 chosen in such a way that their asymptotic regime around the fixed point be independent of $\epsilon$.
 \item Quadratic and higher order terms in the RG invariants were neglected in the scaling functions, i.e.\ $|x|\ll 1$, $|z|\ll 1$,
 and $y\ll 1$. The first of them is automatically fulfilled if $|g_w|\ll 1$, since $g_w$ is an irrelevant field.
 The other two fields are relevant and, therefore, we have the stronger conditions $|g_{m_2}|\ll
 |g_{m_1}|^{\frac{\lambda_{m_2}}{\lambda_{m_1}}}$ and $g_{h^2}\ll |g_{m_1}|^{\frac{\lambda_{h^2}}{\lambda_{m_1}}}$.
 \item Up to this point, we have conditions for the parameters of the effective field theory representing the physical spin
 glass. Translating the above results as a requirement between temperature and magnetic field, we observe that
 $|g_{m_1}|$ is proportional to the reduced temperature, $g_{h^2}\approx w^*h^2\sim H^2$, and $|g_{m_2}|\approx |m_2|\sim H^2$;
 see Eqs.\ (\ref{SK}) and (\ref{F}). 
 As $\lambda_{h^2}$ is the leading relevant eigenvalue, we arrive at
 \[
 H^2\ll |g_{m_1}|^{\frac{\lambda_{h^2}}{\lambda_{m_1}}}\sim \left|\frac{T-T_c}{T_c}\right|^{\frac{\lambda_{h^2}}{\lambda_{m_1}}}\,.
 \]
\end{itemize}
An important consequence of the above analysis is that the ratio $z/y$ is independent of $H^2$ and $z\ll y$: this justifies the simple
three parameter model in \cite{PT} with the fields $|g_{m_1}|$, $g_w$, and $g_{h^2}$ (or equivalently $m_1$, $w_1$, and $h^2$).
Anyway, $z$ entered only the scaling function for $q$ in (\ref{q_hat}).

The scaling functions in Eqs.\ (\ref{q_hat}), (\ref{R_hat}), and (\ref{L_hat}) for $T<T_c$
[and also Eqs.\ (\ref{q_hat_>}), (\ref{R_hat_>}), and (\ref{L_hat_>}) for $T<T_c$]
constitute our basic result: they are
the leading part of a perturbative series, and one could calculate, in principle, any higher order terms in $\epsilon$
and/or in the invariants (say $y$). These series belong completely to the critical fixed point, in other words: they are characteristics
of the zero-magnetic-field fixed point. Their validity is, therefore, independent of the fate of the relevant
couplings (like $h^2$, $m_2$, and $w_i$, $i=2,\dots,5$) under the iteration of the renormalization group, i.e.\ whether
they approach an other fixed point (perturbative or nonperturbative) or flow away to infinity. 

As a matter of fact,
the question is that what information can you extract from these perturbative series. Let's make this point clearer
by the case of the longitudinal mass in (\ref{L_hat}). (For the sake of simplicity, invariants other than $y$
are neglected in the following discussion.) $\hat{\Gamma_L\!\!\!}\,\,\,$ is positive for $y=0$, i.e.\ the zero-field spin glass phase
is longitudinally massive. Although, it is physically plausible that $\Gamma_L$ remains massive in an external field too,
this cannot be verified by (\ref{L_hat}) (or from a longer series), as a nonperturbative zero of $\hat{\Gamma_L\!\!\!}\,\,\,$ is not 
available from such a series.

The situation is fundamentally different for the replicon mass $\hat{\Gamma_R\!\!\!}\,\,\,(y)$ below $T_c$, as it has a {\em perturbative\/}
zero: $y_0=4{w^*}^2+\dots$, whereas the longitudinal mode is massive, $\Gamma_L=2+4\ln 2\,{w^*}^2+\dots$, along this
Almeida-Thouless instability surface. $\hat{\Gamma_R\!\!\!}\,\,\,(y)$ will probably be singular at this zero:
\[
\hat{\Gamma_R\!\!\!}\,\,\,(y)\sim (y-y_0)^{\dot\gamma}\,,
\]
with some exponent. This asymptotic form, however, cannot be verified from the series (\ref{R_hat}) due to the lack
of proper exponentiation. The exponent $\dot\gamma$ cannot be extracted from (\ref{R_hat}), as it does not belong to
the critical fixed point, but possibly to some, at this moment unknown, zero-temperature fixed point. (The scenario
drafted above follows closely the crossover behavior at a bicritical point presented in Ref.\ \cite{Amit_Goldschmidt}.)

\section{Final remarks}\label{conclusion}

It has been shown in the preceding sections how one can detect the critical surface with zero replicon mass (the Almeida-Thouless
critical manifold) asymptotically in the close vicinity of the zero-magnetic-field fixed point {\em perturbatively\/} just
below the upper critical dimension. Nevertheless, this AT critical surface is spanned by relevant scaling fields like
$g_{h^2}$ and $g_{m_2}$, which break the symmetry of the critical zero-magnetic-field fixed point, and runaway RG flows toward infinite
couplings follow \cite{AT2008}. The lack of an attractive perturbative fixed point governing the AT instability surface \cite{BrRo} and
the runaway flows can be understood by the schematic phase diagrams from Refs.\ \cite{PT,Angelini_Biroli}: RG flows along the
AT line terminate into a zero-temperature fixed point, and the effective cubic field theory (fitted to the asymptotics
around the zero-magnetic-field critical transition) is, in fact, {\em not\/} appropriate for representing the
zero-temperature spin glass. A field theory for the low-temperature spin glass is obviously sorely needed for the
understanding of the critical asymptotics along the AT line.

What is claimed above, namely that the existence of a spin glass transition in an external magnetic field may be possible
even if the RG trajectories run away from the critical fixed point without terminating into a perturbative novel fixed point,
has been demonstrated in a simpler model where the interaction depends on the hierarchical distance between the Ising spins:
i.e.\ in the Hierarchical Edwards-Anderson (HEA) model. A first order RG analysis of the generic replica symmetric phase%
\footnote{An analogous study for the short-ranged model in Euclidean $d$-dimensional space can be found in Ref.\ \cite{Iveta}
where the eigenmodes of the linearized RG around the critical fixed point are also presented.} 
in Ref.\ \cite{Castellana_Barbieri_2015} found no relevant fixed point governing the transition in a field:
the couplings  renormalize toward infinite values. Notwithstanding that, a careful Monte Carlo simulation on a modified version of
the HEA \cite{Castellana_Parisi_2015} provided evidence for a transition in nonzero external field by a study of the spin glass susceptibility and the
correlation length associated with it.
Most importantly, Ref.\ \cite{Castellana_Parisi_2015} found transition in nonzero field in the non-mean-field region $\sigma\gtrsim 2/3$
where $\sigma$, the parameter of the HEA analogous to the spacial dimension $d$ of the short-ranged model in Euclidean space, was
within 2\% from the upper critical value $\sigma =2/3$. This clearly shows that the AT instability persists while traversing the analogue
of the upper critical dimension from the mean field to the non-mean-field region, inspite of the absence of a perturbative fixed point
governing the AT critical surface.\cite{Castellana_Barbieri_2015}
One point is still lacking here, namely the observation of the transition perturbatively by computing
the asymptotical behavior of the spin glass susceptibility (or, equivalently, the replicon mass) around the critical fixed point.
This is left to a subsequent work.

As for the short-ranged model, it has been advocated  for some time past\cite{Moore_Bray_2011,Moore_2012} that the lower critical
dimension for the AT line should be $d=6$, i.e.\ that the spin glass transition in an external field disappears just at the upper
critical dimension of the zero-field model. The fault in the arguments of Ref.\ \cite{Moore_Bray_2011} about the behavior 
of the AT line (computed perturbatively for $d>6$), namely that it disappears while approaching $d=6$ from above, was pointed out in \cite{PT}. The issue
was reconsidered in Ref.\ \cite{Yeo_Moore_2015}, admitting now that the six-dimensional AT line cannot be derived by a limiting
process from the perturbative result in $d>6$. Yeo and Moore\cite{Yeo_Moore_2015}, however, incorrectly claimed that the calculation of the
six-dimensional AT line in \cite{PT} was performed by just this wrong limiting process. In fact, the $d=6$ case was studied separately
in Ref.\ \cite{PT},
as it must be, by the special one-loop perturbative RG at the upper critical dimension where the scaling exponent of the cubic
coupling constant is zero. (See also Refs.\ \cite{comment,reply}.) 

From the discussion of the last two paragraphs, and also from the results of the present paper, it follows that the lower critical dimension
for the spin glass transition of the Ising spin glass in an external magnetic field is probably less than $d=6$. One must, however,
emphasize that the perturbative RG is not able to make predictions about the existence of the AT line far below $d=6$.
Numerical simulation results in $d=3$ and $d=4$ (or in the corresponding long-ranged one-dimensional model as a ``proxy'' for the short-ranged
system) in this regard are controversial; see \cite{Janus_2012_1,Larson_et_al_2013} and references therein.

Finally some notes about the perturbative method: The calculations of physical quantities were performed in the present paper
by the combined use of the renormalization
group and a series expansion in terms of the coupling constants. This method is absolutely conventional {\em around\/} a perturbative
fixed point: a perturbative result like (\ref{q_pert}), for instance, is interpreted by the RG ansatz in (\ref{scaling_q}), and the scaling function can be
identified as in (\ref{q_hat}). In the meantime, a consistency check is available by the proper exponentiation of the logarithms of
the temperature-like scaling field. Two peculiarities, however, occur: The first one is due to the quadratic symmetry breaking
caused by the nonzero RS order parameter which leads to the two distinct free propagators, with the replicon mode almost
massless in a small magnetic field below $T_c$. The other one is related
to the replicated natur of the field theory which may cause problems in the $n\to 0$
spin glass limit. Although this limit proved to be quite smooth in our model with the four scaling fields,
the behavior and physical meaning of the remaining modes, like the third mass mode for instance, are not clear.

\appendix
\section{Summary of some one-loop results for the generic replica symmetric theory}

In this Appendix, we provide results which are equally valid in the high and low temperature regimes, assuming that the proper value of
$R$ and $L$, see Eqs.\ (\ref{>}) and  (\ref{<}), must be inserted. 

 \subsection{The equation of state:}

 The order parameter $q$ satisties the implicit equation
 \begin{equation}\label{q}
 2 w^*q=h^2\,q^{-1}-2m_1+2m_2+\left[ -2(w_1-w^*)+w_2-w_3+w_5\right]\,q+q^{-1}\,\frac{1}{N}\sum_{\vec p}\,Y(p)\,\,,
 \end{equation}
 with the one-loop graph
 \begin{multline*}
  Y(p)=\left(w_2+\frac{1}{3}\,w_3+\frac{4}{3}\,w_4-\frac{1}{3}\,w_5-w_6-\frac{4}{3}\,w_7\right)\,\bar G_R\\[8pt]
  +\left(3w_1-2w_2+\frac{4}{3}\,w_3+\frac{4}{3}\,w_4-\frac{1}{3}\,w_5-w_6-\frac{4}{3}\,w_7\right)\, 2\bar{m_2}\,\bar G_R\bar G_L
  +\left(4w_1-2w_2+2w_3-2w_5-2w_6\right)\,(-\bar{m_2}+2\bar{m_3})\,{\bar G_L}^2.
 \end{multline*}
This one-loop integral can be computed, and one gets
\begin{multline}\label{Y}
 w^*\,\frac{1}{N}\sum_{\vec p}\,Y(p)={w^*}^2\, |g_{m_1}|^2 \,\times\\[5pt]
 (1-2x)^{-1/2}\bigg[\dfrac{1}{2}(R-L)\,|g_{m_1}|^{-1}+\dfrac{1}{2}(L-R)(L-3R)\,\ln |g_{m_1}|
 +\dfrac{3}{2}R^2\ln R+\dfrac{1}{2}L(L-4R)\ln L+L(L-R)\bigg] +O({w^*}^4).
\end{multline}

\subsection{The replicon mass}

The one-loop formula for the replicon mass has been published in \cite{rscikk}; see Eqs.\ (49a-h) and (62). Here we reproduce it
in terms of the set of bare parameters used throughout the present paper and for $n=0$:
\begin{equation}\label{Gamma_R}
 \Gamma_R=2m_1+2 w^*q+2\left[ (w_1-w^*)-w_2+\frac{1}{3}\,w_3\right]\,q-\frac{1}{N}\sum_{\vec p}\,\sigma_R\,
\end{equation}
with the replicon self energy:
\begin{multline*}
\sigma_R=\Big(-2w_1^2-\frac{4}{3}\,w_1w_3-\frac{16}{3}\,w_1w_4- \frac{8}{3}\,w_1w_5+2w_2^2+ \frac{8}{3}\,w_2w_3
+\frac{16}{3}\,w_2w_4+\frac{4}{3}\,w_2w_5+\frac{2}{9}\,w_3^2-\frac{16}{9}\,w_3w_4
-\frac{8}{9}\,w_3w_5\\[8pt]+\frac{4}{9}\,w_5^2\Big)\times\bar G_R^2
+\Big(-2w_1^2+12w_1w_2+\frac{4}{3}\,w_1w_3-\frac{16}{3}\,w_1w_4- \frac{16}{3}\,w_1w_5-8w_2^2+ \frac{4}{3}\,w_2w_3
+\frac{16}{3}\,w_2w_4+\frac{8}{3}\,w_2w_5
+\frac{6}{9}\,w_3^2\\[8pt]-\frac{16}{9}\,w_3w_4
-\frac{16}{9}\,w_3w_5+\frac{8}{9}\,w_5^2\Big)\times 2\bar{m_2}\bar G_R^2\bar G_L
+\Big(-8w_1^2+16w_1w_2-\frac{16}{3}\,w_1w_3-8w_2^2+ \frac{16}{3}\,w_2w_3
-\frac{8}{9}\,w_3^2\Big)
\times
(-\bar{m_2}+2\bar{m_3})\,\bar G_R\bar G_L^2\\[8pt]
+\frac{1}{9}\,(6w_1-3w_2+2w_3-2w_5)^2
\times 4\bar{m_2}^2\,\bar G_R^2\bar G_L^2\,.
\end{multline*}
Performing the momentum integral provides
\begin{multline}\label{sigma_R}
\frac{1}{N}\sum_{\vec p}\,\sigma_R={w^*}^2\, |g_{m_1}| \,\times\\[5pt]
 (1-2x)^{-1}\bigg[-|g_{m_1}|^{-1}+(L-3R)\,\ln |g_{m_1}|
 +\dfrac{R(4L+3R)}{L-R}\ln R+\dfrac{L(L-8R)}{L-R}\ln L+2(R+2L)\bigg] +O({w^*}^4).
\end{multline}

\subsection{The longitudinal mass}
 
 The first order expression for the longitudinal mass takes the form (for $n=0$)
 \begin{equation}\label{Gamma_L}
 \Gamma_L=2m_1-2m_2+4 w^*q+2\,\left[ 2(w_1-w^*)-w_2+w_3-w_5\right]\,q-\frac{1}{N}\sum_{\vec p}\,\sigma_L\,
\end{equation}
with the longitudinal self energy (which is identical with the anomalous one when $n=0$); see Eqs.\ (49a-h), (63), and (64) of
\cite{rscikk}:
\begin{multline*}
 \sigma_L=
\left[6\,(w_1-w_2+\frac{1}{3}\,w_3)^2-\frac{4}{3}\,(2w_1-w_2+w_3-w_5-w_6)\,(3w_1-3w_2+w_3+4w_4+2w_5-4w_7)\right]
\times\bar G_R^2\\[10pt]
-\frac{8}{3}\,(2w_1-w_2+w_3-w_5-w_6)\,(3w_1-3w_2+w_3+4w_4+2w_5-4w_7)
\times (2\bar{m_2}\,\bar G_R^2\bar G_L+2\bar{m_2}^2\,\bar G_R^2\bar G_L^2)\\[10pt]
-8(-2w_1+w_2-w_3+w_5+w_6)^2\times(-\bar{m_2}+2\bar{m_3})\,\bar G_L^3\,\,.
\end{multline*}
After integration it becomes:
\begin{multline}\label{sigma_L}
\frac{1}{N}\sum_{\vec p}\,\sigma_L={w^*}^2\, |g_{m_1}| \,\times
 (1-2x)^{-1}\bigg[-|g_{m_1}|^{-1}-2R\,\ln |g_{m_1}|
 +6R\ln R-8R\ln L+(8L-9R)\bigg] +O({w^*}^4).
\end{multline}


\begin{thebibliography}{18}

\expandafter\ifx\csname natexlab\endcsname\relax\def\natexlab#1{#1}\fi
\expandafter\ifx\csname bibnamefont\endcsname\relax
  \def\bibnamefont#1{#1}\fi
\expandafter\ifx\csname bibfnamefont\endcsname\relax
  \def\bibfnamefont#1{#1}\fi
\expandafter\ifx\csname citenamefont\endcsname\relax
  \def\citenamefont#1{#1}\fi
\expandafter\ifx\csname url\endcsname\relax
  \def\url#1{\texttt{#1}}\fi
\expandafter\ifx\csname urlprefix\endcsname\relax\def\urlprefix{URL }\fi
\providecommand{\bibinfo}[2]{#2}
\providecommand{\eprint}[2][]{\url{#2}}

\bibitem[{\citenamefont{Wilson and Kogut}(1974)}]{Wilson_Kogut}
\bibinfo{author}{\bibfnamefont{K.}~\bibnamefont{Wilson}} \bibnamefont{and}
  \bibinfo{author}{\bibfnamefont{J.}~\bibnamefont{Kogut}},
  \bibinfo{journal}{Physics Reports} \textbf{\bibinfo{volume}{12}},
  \bibinfo{pages}{75} (\bibinfo{year}{1974}).

\bibitem[{\citenamefont{Edwards and Anderson}(1975)}]{EA}
\bibinfo{author}{\bibfnamefont{S.~F.} \bibnamefont{Edwards}} \bibnamefont{and}
  \bibinfo{author}{\bibfnamefont{P.~W.} \bibnamefont{Anderson}},
  \bibinfo{journal}{J. Phys. F} \textbf{\bibinfo{volume}{5}},
  \bibinfo{pages}{965} (\bibinfo{year}{1975}).
 
\bibitem[{\citenamefont{Harris et~al.}(1976)\citenamefont{Harris, Lubensky, and
  Chen}}]{HaLuCh76}
\bibinfo{author}{\bibfnamefont{A.~B.} \bibnamefont{Harris}},
  \bibinfo{author}{\bibfnamefont{T.~C.} \bibnamefont{Lubensky}},
  \bibnamefont{and} \bibinfo{author}{\bibfnamefont{J.-H.} \bibnamefont{Chen}},
  \bibinfo{journal}{Phys. Rev. Lett.} \textbf{\bibinfo{volume}{36}},
  \bibinfo{pages}{415} (\bibinfo{year}{1976}).

\bibitem[{\citenamefont{Pytte and Rudnick}(1979)}]{PytteRudnick79}
\bibinfo{author}{\bibfnamefont{E.}~\bibnamefont{Pytte}} \bibnamefont{and}
  \bibinfo{author}{\bibfnamefont{J.}~\bibnamefont{Rudnick}},
  \bibinfo{journal}{Phys. Rev. B} \textbf{\bibinfo{volume}{19}},
  \bibinfo{pages}{3603} (\bibinfo{year}{1979}).

\bibitem[{\citenamefont{Bray and Moore}(1979)}]{BrMo79}
\bibinfo{author}{\bibfnamefont{A.~J.} \bibnamefont{Bray}} \bibnamefont{and}
  \bibinfo{author}{\bibfnamefont{M.~A.} \bibnamefont{Moore}},
  \bibinfo{journal}{J. Phys. C} \textbf{\bibinfo{volume}{12}},
  \bibinfo{pages}{79} (\bibinfo{year}{1979}).
 
\bibitem[{\citenamefont{Temesv{\'a}ri et~al.}(2002)\citenamefont{Temesv{\'a}ri,
  De~Dominicis, and Pimentel}}]{rscikk}
\bibinfo{author}{\bibfnamefont{T.}~\bibnamefont{Temesv{\'a}ri}},
  \bibinfo{author}{\bibfnamefont{C.}~\bibnamefont{De~Dominicis}},
  \bibnamefont{and} \bibinfo{author}{\bibfnamefont{I.~R.}
  \bibnamefont{Pimentel}}, \bibinfo{journal}{Eur. Phys. J. B}
  \textbf{\bibinfo{volume}{25}}, \bibinfo{pages}{361} (\bibinfo{year}{2002}),
  \eprint{cond-mat/0202162}.

\bibitem[{\citenamefont{de~Almeida and Thouless}(1978)}]{AT}
\bibinfo{author}{\bibfnamefont{J.~R.~L.} \bibnamefont{de~Almeida}}
  \bibnamefont{and} \bibinfo{author}{\bibfnamefont{D.~J.}
  \bibnamefont{Thouless}}, \bibinfo{journal}{J. Phys. A}
  \textbf{\bibinfo{volume}{11}}, \bibinfo{pages}{983} (\bibinfo{year}{1978}).

\bibitem[{\citenamefont{Green et~al.}(1983)\citenamefont{Green, Moore, and
  Bray}}]{GrMoBr83}
\bibinfo{author}{\bibfnamefont{J.~E.} \bibnamefont{Green}},
  \bibinfo{author}{\bibfnamefont{M.~A.} \bibnamefont{Moore}}, \bibnamefont{and}
  \bibinfo{author}{\bibfnamefont{A.~J.} \bibnamefont{Bray}},
  \bibinfo{journal}{J. Phys. C} \textbf{\bibinfo{volume}{16}},
  \bibinfo{pages}{L815} (\bibinfo{year}{1983}).
 
\bibitem[{\citenamefont{Sherrington and Kirkpatrick}(1975)}]{SK}
\bibinfo{author}{\bibfnamefont{D.}~\bibnamefont{Sherrington}} \bibnamefont{and}
  \bibinfo{author}{\bibfnamefont{S.}~\bibnamefont{Kirkpatrick}},
  \bibinfo{journal}{\prl} \textbf{\bibinfo{volume}{35}}, \bibinfo{pages}{1792}
  (\bibinfo{year}{1975}).

\bibitem[{\citenamefont{Temesv\'ari}(2008)}]{AT2008}
\bibinfo{author}{\bibfnamefont{T.}~\bibnamefont{Temesv\'ari}},
  \bibinfo{journal}{\prb} \textbf{\bibinfo{volume}{78}},
  \bibinfo{pages}{220401(R)} (\bibinfo{year}{2008}), \eprint{arXiv:0809.1839}.

\bibitem[{\citenamefont{Parisi and Temesv\'ari}(2012)}]{PT}
\bibinfo{author}{\bibfnamefont{G.}~\bibnamefont{Parisi}} \bibnamefont{and}
  \bibinfo{author}{\bibfnamefont{T.}~\bibnamefont{Temesv\'ari}},
  \bibinfo{journal}{Nucl.\ Phys.\ B} \textbf{\bibinfo{volume}{858}},
  \bibinfo{pages}{293} (\bibinfo{year}{2012}), \eprint{arXiv:1111.3313}.
  
\bibitem[{\citenamefont{Pimentel et~al.}(2002)\citenamefont{Pimentel,
  Temesv{\'a}ri, and De~Dominicis}}]{Iveta}
\bibinfo{author}{\bibfnamefont{I.~R.} \bibnamefont{Pimentel}},
  \bibinfo{author}{\bibfnamefont{T.}~\bibnamefont{Temesv{\'a}ri}},
  \bibnamefont{and}
  \bibinfo{author}{\bibfnamefont{C.}~\bibnamefont{De~Dominicis}},
  \bibinfo{journal}{\prb} \textbf{\bibinfo{volume}{65}},
  \bibinfo{pages}{224420} (\bibinfo{year}{2002}), \eprint{cond-mat/0204615}.

\bibitem[{\citenamefont{Wegner}(1972)}]{Wegner}
\bibinfo{author}{\bibfnamefont{F.~J.} \bibnamefont{Wegner}},
  \bibinfo{journal}{\prb} \textbf{\bibinfo{volume}{5}}, \bibinfo{pages}{4529}
  (\bibinfo{year}{1972}).

\bibitem[{\citenamefont{Amit and Goldschmidt}(1978)}]{Amit_Goldschmidt}
\bibinfo{author}{\bibfnamefont{D.~J.} \bibnamefont{Amit}} \bibnamefont{and}
  \bibinfo{author}{\bibfnamefont{Y.~Y.} \bibnamefont{Goldschmidt}},
  \bibinfo{journal}{Annals of Physics} \textbf{\bibinfo{volume}{114}},
  \bibinfo{pages}{356} (\bibinfo{year}{1978}).
 
\bibitem[{\citenamefont{Bray and Roberts}(1980)}]{BrRo}
\bibinfo{author}{\bibfnamefont{A.~J.} \bibnamefont{Bray}} \bibnamefont{and}
  \bibinfo{author}{\bibfnamefont{S.~A.} \bibnamefont{Roberts}},
  \bibinfo{journal}{J. Phys. C} \textbf{\bibinfo{volume}{13}},
  \bibinfo{pages}{5405} (\bibinfo{year}{1980}).

\bibitem[{\citenamefont{Angelini and Biroli}(2015)}]{Angelini_Biroli}
\bibinfo{author}{\bibfnamefont{M.~C.} \bibnamefont{Angelini}} \bibnamefont{and}
  \bibinfo{author}{\bibfnamefont{G.}~\bibnamefont{Biroli}},
  \bibinfo{journal}{Phys. Rev. Lett.} \textbf{\bibinfo{volume}{114}},
  \bibinfo{pages}{095701} (\bibinfo{year}{2015}).

\bibitem[{\citenamefont{Castellana and
  Barbieri}(2015)}]{Castellana_Barbieri_2015}
\bibinfo{author}{\bibfnamefont{M.}~\bibnamefont{Castellana}} \bibnamefont{and}
  \bibinfo{author}{\bibfnamefont{C.}~\bibnamefont{Barbieri}},
  \bibinfo{journal}{\prb} \textbf{\bibinfo{volume}{91}},
  \bibinfo{pages}{024202} (\bibinfo{year}{2015}), \eprint{arXiv:1503.00066}.
  
\bibitem[{\citenamefont{Castellana and Parisi}(2015)}]{Castellana_Parisi_2015}
\bibinfo{author}{\bibfnamefont{M.}~\bibnamefont{Castellana}} \bibnamefont{and}
  \bibinfo{author}{\bibfnamefont{G.}~\bibnamefont{Parisi}},
  \bibinfo{journal}{Scientific Reports} \textbf{\bibinfo{volume}{5}},
  \bibinfo{pages}{8697} (\bibinfo{year}{2015}), \eprint{arXiv:1503.02103}.

\bibitem[{\citenamefont{Moore and Bray}(2011)}]{Moore_Bray_2011}
\bibinfo{author}{\bibfnamefont{M.}~\bibnamefont{Moore}} \bibnamefont{and}
  \bibinfo{author}{\bibfnamefont{A.}~\bibnamefont{Bray}},
  \bibinfo{journal}{\prb} \textbf{\bibinfo{volume}{83}},
  \bibinfo{pages}{224408} (\bibinfo{year}{2011}), \eprint{arXiv:1102.1675}.

\bibitem[{\citenamefont{Moore}(2012)}]{Moore_2012}
\bibinfo{author}{\bibfnamefont{M.}~\bibnamefont{Moore}},
  \bibinfo{journal}{\pre} \textbf{\bibinfo{volume}{86}},
  \bibinfo{pages}{031114} (\bibinfo{year}{2012}), \eprint{arXiv:1206.4492}.

\bibitem[{\citenamefont{Yeo and Moore}(2015)}]{Yeo_Moore_2015}
\bibinfo{author}{\bibfnamefont{J.}~\bibnamefont{Yeo}} \bibnamefont{and}
  \bibinfo{author}{\bibfnamefont{M.}~\bibnamefont{Moore}},
  \bibinfo{journal}{\prb} \textbf{\bibinfo{volume}{91}},
  \bibinfo{pages}{104432} (\bibinfo{year}{2015}), \eprint{arXiv:1412.2448}.
 
\bibitem[{\citenamefont{Temesv\'ari}(2016)}]{comment}
\bibinfo{author}{\bibfnamefont{T.}~\bibnamefont{Temesv\'ari}},
  \bibinfo{journal}{\prb} \textbf{\bibinfo{volume}{94}},
  \bibinfo{pages}{176401} (\bibinfo{year}{2016}), \eprint{arXiv:1610.04747}.

\bibitem[{\citenamefont{Yeo and Moore}(2016)}]{reply}
\bibinfo{author}{\bibfnamefont{J.}~\bibnamefont{Yeo}} \bibnamefont{and}
  \bibinfo{author}{\bibfnamefont{M.}~\bibnamefont{Moore}},
  \bibinfo{journal}{\prb} \textbf{\bibinfo{volume}{94}},
  \bibinfo{pages}{176402} (\bibinfo{year}{2016}), \eprint{arXiv:1610.05764}.

\bibitem[{\citenamefont{Banos et~al.}(2012)\citenamefont{Banos, Cruz,
  Fernandez, Gil-Narvion, Gordillo-Guerrero, Guidetti, Iniguez, Maiorano, ,
  Marinari et~al.}}]{Janus_2012_1}
\bibinfo{author}{\bibfnamefont{R.~A.} \bibnamefont{Banos}},
  \bibinfo{author}{\bibfnamefont{A.}~\bibnamefont{Cruz}},
  \bibinfo{author}{\bibfnamefont{L.}~\bibnamefont{Fernandez}},
  \bibinfo{author}{\bibfnamefont{J.}~\bibnamefont{Gil-Narvion}},
  \bibinfo{author}{\bibfnamefont{A.}~\bibnamefont{Gordillo-Guerrero}},
  \bibinfo{author}{\bibfnamefont{M.}~\bibnamefont{Guidetti}},
  \bibinfo{author}{\bibfnamefont{D.}~\bibnamefont{Iniguez}},
  \bibinfo{author}{\bibfnamefont{A.}~\bibnamefont{Maiorano}}, ,
  \bibinfo{author}{\bibfnamefont{E.}~\bibnamefont{Marinari}},
  \bibnamefont{et~al.}, \bibinfo{journal}{PNAS} \textbf{\bibinfo{volume}{109}},
  \bibinfo{pages}{6452} (\bibinfo{year}{2012}), \bibinfo{note}{(Janus
  Collaboration)}, \eprint{arXiv:1202.5593}.
  
\bibitem[{\citenamefont{Larson et~al.}(2013)\citenamefont{Larson, Katzgraber,
  Moore, and Young}}]{Larson_et_al_2013}
\bibinfo{author}{\bibfnamefont{D.}~\bibnamefont{Larson}},
  \bibinfo{author}{\bibfnamefont{H.~G.} \bibnamefont{Katzgraber}},
  \bibinfo{author}{\bibfnamefont{M.}~\bibnamefont{Moore}}, \bibnamefont{and}
  \bibinfo{author}{\bibfnamefont{A.}~\bibnamefont{Young}},
  \bibinfo{journal}{\prb} \textbf{\bibinfo{volume}{87}},
  \bibinfo{pages}{024414} (\bibinfo{year}{2013}), \eprint{arXiv:1211.7297}.

\end{thebibliography}

\end{document}